\documentclass{article}

\usepackage[left=2.5cm,right=2.5cm,top=2.75cm,bottom=2.75cm]{geometry}
\usepackage{microtype}
\usepackage{graphicx}
\usepackage{subfigure}
\usepackage{amsmath}
\usepackage{amssymb}
\usepackage{bm}
\usepackage{soul}
\usepackage{booktabs} 
\usepackage{color,colortbl}
\usepackage[dvipsnames]{xcolor}
\usepackage{cite}

\usepackage{hyperref}

\usepackage{relsize}

\usepackage{stackengine}
\usepackage{scalerel}
\def\rlwd{.4pt}
\def\rlht{1.1pt}
\def\shatvrule{\rule{\rlwd}{\rlht}}
\def\shat#1{%
 \ThisStyle{%
  \setbox0=\hbox{$\SavedStyle#1$}%
  \stackon[0pt]{\stackon[1pt]{\ensuremath{\SavedStyle#1}}{%
    \shatvrule\kern\wd0\kern-\rlwd\kern-\rlwd\shatvrule}}%
    {\rule{\wd0}{\rlwd}}%
 }%
}

\title{\textbf{Burst-dependent plasticity and dendritic amplification support target-based learning and hierarchical imitation learning}}

\author{
Cristiano Capone$^\ast$\\
\normalsize INFN, Sezione di Roma, Rome, Italy\\
\textsf{cristiano0capone@gmail.com}\\
\and
Cosimo Lupo$^\ast$\\
\normalsize INFN, Sezione di Roma, Rome, Italy\\
\and
Paolo Muratore\\
\normalsize SISSA, International School for\\
\normalsize Advanced Studies, Trieste, Italy\\
\and
Pier Stanislao Paolucci\\
\normalsize INFN, Sezione di Roma, Rome, Italy\\
}

\date{}

\definecolor{cristianoorange}{rgb}{0.9,0.3,0.}
\definecolor{minoblue}{rgb}{0.,0,0.8}
\definecolor{minogreen}{rgb}{0.565,0.933,0.565}
\definecolor{paoloviolet}{rgb}{0.788, 0.36, 0.92}
\definecolor{pierred}{rgb}{0.9, 0.1, 0.1}

\newcommand{\lock}[0]{
\mbox{\scriptsize$\begin{smallmatrix}\mathsmaller{\bm{\cap}} \\[-1pt] \mathlarger{\blacksquare} \end{smallmatrix}$}
}

\begin{document}

\twocolumn[
\maketitle
\begin{center}
\vspace{-0.7cm}
\normalsize $^\ast$These authors contributed equally to this work
\end{center}

\vspace{0.3cm}

\vskip 0.1in

\begin{center}
\begin{abstract}
The brain can learn to solve a wide range of tasks with high temporal and energetic efficiency.
However, most biological models are composed of simple single compartment neurons and cannot achieve the state-of-art performances of artificial intelligence.
We propose a multi-compartment model of pyramidal neuron, in which bursts and dendritic input segregation give the possibility to plausibly support a biological target-based learning. In target-based learning, the internal solution of a problem (a spatio temporal pattern of bursts in our case) is suggested to the network, bypassing the problems of error backpropagation and credit assignment.
Finally, we show that this neuronal architecture naturally supports the orchestration of “hierarchical imitation learning”, enabling the decomposition of challenging long-horizon decision-making tasks into simpler subtasks.
\end{abstract}
\end{center}
\vskip 0.3in
]

\section{Introduction}

The brain can learn a wide range of tasks very efficiently in terms of energy consumption and required evidences, motivating the search for biologically inspired learning rules for improving the efficiency of artificial intelligence.
Most biologically plausible neural networks are composed so far of point neurons. Despite recent outstanding advances in this field \cite{nicola2017supervised,bellec2020}, biologically plausible neural networks cannot achieve the state-of-art performances of artificial intelligence (e.g. they struggle to solve the credit assignment problem \cite{payeur2021burst}).

Recent findings on dendritic computational properties \cite{poirazi2020illuminating} and on the complexity of pyramidal neurons dynamics \cite{larkum2013cellular} motivated the study of multi-compartment neuron model in the development of new biologically plausible learning rules \cite{urbanczik2014learning,guerguiev2017towards,sacramento2018dendritic,payeur2021burst}.

Recent works have proposed that segregation of dendritic input (neurons receive sensory information and higher-order feedback in segregated compartments) \cite{guerguiev2017towards} and generation of high-frequency bursts of spikes \cite{payeur2021burst} would support backpropagation in biological neurons.
However, these approaches require propagating errors with a fine spatio-temporal structure to all the neurons. It is not clear whether this is possible in biological networks. For this reason, in the last few years, target-based approaches \cite{lee2015difference,depasquale2018full,manchev2020target,meulemans2020theoretical,muratore2021target} started to gain more and more interest.

In a target-based learning framework, the targets, rather than the errors, are propagated through the network \cite{lee2015difference,manchev2020target}. In this framework, it is possible to directly suggest to the network the internal solution to a task \cite{depasquale2018full,muratore2021target,capone2021error}.
However, target-based approaches require evaluating at the same time the spontaneous activity and the target activity of the network \cite{depasquale2018full,muratore2021target}. This is usually solved by evaluating the two activities in two different networks, which is not natural in terms of biological plausibility. 

In the present work, we show that bursts and dendritic input segregation offer a natural solution to this dilemma.
In our model, pyramidal neurons rely on two different apical dendritic compartments to simultaneously evaluate the target and the spontaneous activity. A coincidence mechanism between basal and apical inputs generating the burst \cite{larkum2013cellular} eventually defines the (target or spontaneous) spatio-temporal bursting dynamics of the network.

We exploit dendritic computation in our model, to let abstract signals act as teaching signals which drive the learning procedure in a biologically plausible fashion.

Finally, we show that this neuronal architecture naturally allows for orchestrating “hierarchical imitation learning”, enabling the decomposition of challenging long-horizon decision-making tasks into simpler subtasks \cite{le2018hierarchical,pateria2021hierarchical}.

\section{Results}

\subsection{Target-based learning with bursts}

\begin{figure*}
\centering
\includegraphics[width=160mm]{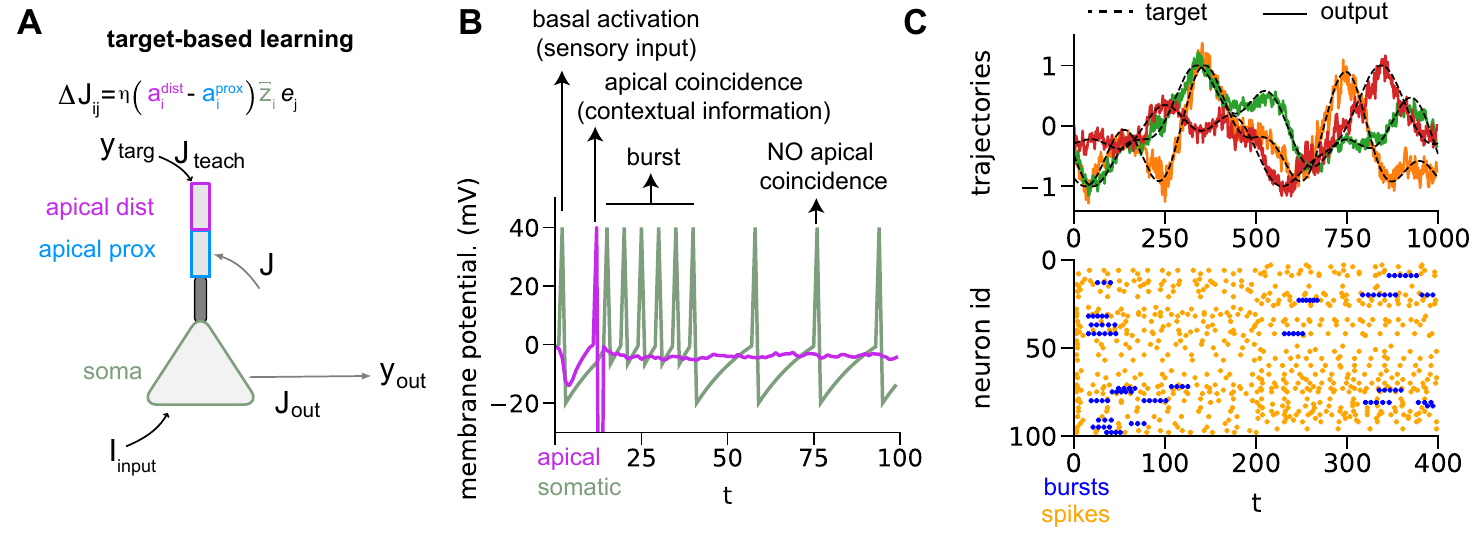}
\caption{
\textbf{Model structure} 
\textbf{A}. The model of a pyramidal neuron, consisting of two separated compartments, the basal and the apical ones. The latter is further divided into two regions, proximal (receiving recurrent connections from the network) and distal (receiving teaching/context signals from other areas of the cortex).
\textbf{B}. In addition to isolated spike signals emitted by the soma, a coincidence mechanism between basal and apical compartments allows for the generation of high-frequency bursts of spikes.
\textbf{C}. Store-and-recall of a 3D trajectory. The target output is automatically encoded into a spatio-temporal pattern of bursts (bottom panel), learned online thanks to the plasticity of recurrent connections, allowing for reliable reproduction of the target trajectory (top panel).
}
\label{fig1}
\end{figure*}

We define a model of pyramidal neuron (Fig.\ref{fig1}A, bottom) composed of three separated compartments, the basal one (i.e. the soma, receiving the sensorial input), and two apical ones, the proximal apical compartment (receiving recurrent connections from the network) and the distal apical compartment (receiving the context/teaching signal from other areas of the cortex, with a higher level of abstraction). 

The spike emitted by the soma is described by variable $z_i^t$, which is equal to $one$ when the spike is emitted at time $t$ and $zero$ otherwise. The spikes emitted by the proximal and distal apical compartments are described by the variables $a_i^t$ and $a_i^{\star \, t}$, respectively. The underlying idea is that the distal compartment provides a target for the proximal one, motivating the use of the superscript symbol $\star$, which indicates the variables concerning the targets.

In addition, following \cite{larkum2013cellular} a coincidence mechanism between the basal and the apical compartments has been implemented, yielding high-frequency bursts of spikes. In more detail, after a somatic spike, $z_i^t=1$, a coincidence window is opened for a time interval $\Delta T$. This is described by the variable $\shat{z}_i^t$, the indicator function for $t' \in [t, t + \Delta T]$, which is $one$ during this time window. If a spike is generated by the distal or proximal apical compartments within such time window, $a_i^{t'}=1$ or $a_i^{\star \, t'}=1$ with $t'\in[t_k, t_k + \Delta T]$, a high frequency burst of spikes is then produced (Fig.\ref{fig1}B).
The proximal and distal bursts can be respectively defined as
\begin{align*}
B_{i}^{t+1} &= \shat{z}_i^{t}  \, a_{i}^{t+1}\\
B_{i}^{\star, t+1} &= \shat{z}_i^{t} \, a_{i}^{\star, t+1}
\end{align*}
This architecture supports a burst-dependent learning rule (Fig.\ref{fig1}A, top), enabling target-based learning. More specifically, the pattern of bursts defined by the proximal compartment (receiving the recurrent connections from the network) should mimic the ones defined by the distal compartment (which receives the teaching signal). This is possible by using the following plasticity rule for recurrent weights $J_{ij}^{b \to p}$ (which can be derived analytically through a likelihood maximization, see methods for details):
\begin{equation}
\Delta J_{ij}^{b \to p} = \eta   \left[a_{i}^{\star, t+1} - a_{i}^{t+1} \right] \shat{z}_{i}^t \, e_j^t
\label{plasticity_rule}
\end{equation}
where $e_j^t = \partial u_i^t/\partial J_{ij}^{b \to p}$ is referred to in the literature as the spike response function \cite{urbanczik2014learning}.

Intuitively, such plasticity rule aims at aligning in time apical proximal spikes with apical distal ones when the somatic window $\shat{z}_i^t$ is open. We remark that such learning rule can be computed online, and requires only observables which are locally accessible to the synapses in space and time.

As a first learning instance, we propose the store-and-recall of a 3D trajectory $y_k^{\star \ t}$ ($k = 1,\dots,3$, $t = 1,\dots,T$, $T=1000$) in a network of $N = 500$ neurons ($400$ excitatory plus $100$ inhibitory). We chose $y_k^{\star \ t}$ as a temporal pattern composed of $3$ independent continuous signals, each of which specified as the superposition of the four frequencies $f \in \left\{1, 2, 3, 5 \right\}$ Hz with uniformly extracted random amplitude $A \in \left[0.5, 2.0 \right]$, and phases $\phi \in \left[0, 2 \pi \right]$:
\begin{equation*}
y_k^{\star \ t} = \sum_{n=1}^4 A_{k,n}\cos{(2\pi f_{k,n} t + \phi_{k,n})} \quad , \qquad k=1,2,3
\end{equation*}
This trajectory is randomly projected through a Gaussian matrix with variance $\sigma_{\mathrm{targ}}^2$ to the apical (distal) dendrites of the network as a teaching signal. This input shapes the spatio-temporal pattern of spikes $a_{i}^{\star, t}$ from the distal apical compartment, as well as the target spatio-temporal pattern of bursts $B_i^{\star \,t}$ (Fig.\ref{fig1}C bottom, blue points) as described above.

A clock signal serving as a sensorial input is randomly projected (through a gaussian matrix with variance $\sigma_{\mathrm{in}}^2$) to the somatic dendrites. In more detail, the clock is here modeled as a sort of time step function with $I$  steps, such that at each time $t$ only component $i=\lfloor I\cdot t/T\rfloor$ is equal to one, while others are zero (see Table.\ref{table1} for model parameters).

Learning is numerically implemented by several presentations of the same target trajectory $y^{\star}$ to the distal apical compartments, each time adjusting recurrent weights $J_{ij}^{b \to p}$ according to \eqref{plasticity_rule}.

Bursting internal activity, which represents the actual quantity mimicking the target, is translated into the output $y$ by means of a read-out matrix $J_{\mathrm{out}}$, randomly initialized and to be trained following the rule derived by minimizing the mean squared error between the target output and the network's output:
\begin{equation}
    \Delta J^{\mathrm{out}}_{ki} = \eta_{\mathrm{out}} \left[y^{\star \, t}_k - \sum_{h} J^{\mathrm{out}}_{kh} \hat{B}^t_h\right] \hat{B}^t_i
\end{equation}
where $\hat{B}_i^t$ is a time-smoothed version of burst variable $B$ (see methods for details).

At the end of the learning the plasticity of recurrent connections allows for a reliable reproduction of the target 3D trajectory (see Fig.\ref{fig1}C, top, $mse=0.01$), with an internal bursting activity reproducing the target one (Fig.\ref{fig1}C, bottom).

\subsection{Apical signals as a flexible context selection}

\begin{figure*}[h]
\centering
\includegraphics[width=170mm]{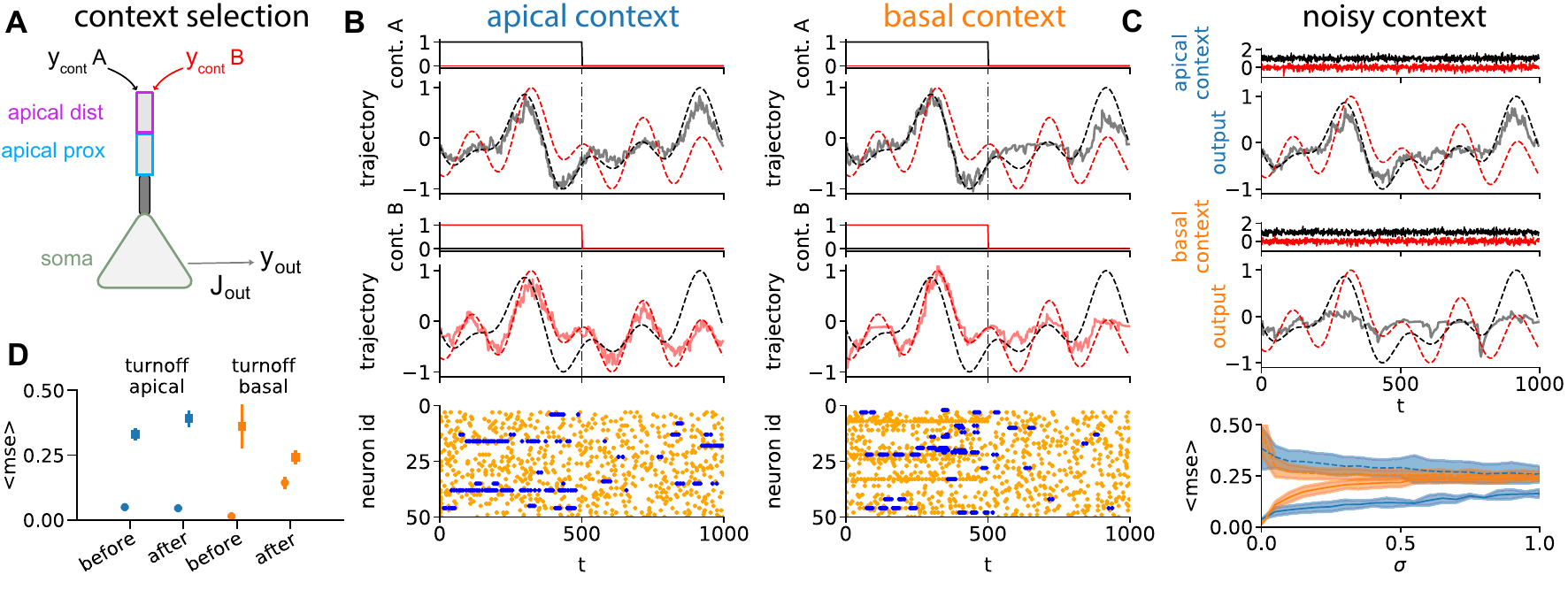}
\caption{
\textbf{Apical signals for dynamics selection.}
\textbf{A}. Model of pyramidal neuron where a binary context signal  (A or B) is projected on the apical distal compartment. The target to be reproduced by the network changes according to which context is active.
\textbf{B}. The network is able to reproduce the correct output trajectory even if the context is provided only in the first time steps. An alternative model in which the context is projected on the basal compartment is no longer able to reproduce the correct output trajectory.
\textbf{C}. (top) The trajectory produced by the network, in presence of noisy apical context A ($\sigma = 0.2$ black solid line) is similar to the trajecory targeted by the context A (black dashed line) and different from the trajectory targeted by the context B. Inset, the noisy context signal (red dashed line). (middle) The trajectory produced by the network, in presence of noisy basal context A ($\sigma = 0.2$ black line) is NOT similar to the trajectory targeted by the context A (black dashed line). Inset, the noisy context signal. (bottom) The average performances of the apical/basal (blue/orange) context as a function of the noise standard deviation $\sigma$. Solid lines: mse between the output and the target output. Dashed lines: mse between the output and the trajectory targeted by the other context. Averages and error bars are intended over many independent network/target realizations. 
\textbf{D}. Summary of performances of the two model versions (context projected on apical vs basal compartment) during ``turnoff'' test in the middle of the trajectory. Mean square error in the second part of the trajectory (no context) is compared with respect to error in the first part (context still active); mean and variance are intended over many independent network/target realizations.
}
\label{fig2}
\end{figure*}

In this section, we show that it is possible to project to the distal apical compartment signal context (through a random matrix with variance $\sigma_\mathrm{cont}^2$ ) to flexibly select and recall one of the trajectory stored in the network.

In the simplest configuration, two different context binary signals can be projected on the apical compartment, A or B (Fig.\ref{fig2}A). In detail, context signal is modeled as a 2D signal, which is $C^t = (1,0)$ for the context A and $C^t = (0,1)$ for the context B.

During the training, each context is associated with a well defined target to learn (again a 3D trajectory, as defined in the previous section, Fig.\ref{fig2}B, left side, in red and black respectively, only one of the three trajectories is reported for simplicity). To stabilize the learning, we exploited the trick of halving the learning rates $\eta$ and $\eta_{\mathrm{out}}$ every $100$ training iterations. The orthogonality of the contexts and related targets is stressed by imposing a sparsification (of $75\%$ in the present case) in the random matrices we use to project the context and the target on the apical compartments of the network.

During the recall phase, the teacher signal is no longer present, while the context signal suggests to the network which of the learned trajectory to reproduce.
We show that when the context is projected to the network, the desired output is correctly recalled (Fig.\ref{fig2}B, left side). Moreover, if the context signal is turned off in the middle of the trajectory, the network is able to self-sustain its inner dynamics, thanks to recurrent connections (Fig.\ref{fig2}B, left side), and correctly replicate the selected trajectory.

The context is here a ``suggestion'', so that once started the reproduction of the correct output trajectory, the context itself becomes useless. 


To demonstrate the importance to project the context signal in the apical compartments we compare these results with the case in which the context is projected in the basal ones (both during the training and the retrieval phases).

In this case, the desired trajectory is correctly retrieved when the contect is on (Fig.\ref{fig2}B, right side).

However, we observe that the basal context is interpreted as a necessary input, so that after the turn-off the network is no longer able to sustain bursts creation, in turn causing a dramatic drop in the test performances (Fig.\ref{fig2}B, right side). Average mean square errors, measured against both the correct target trajectory and the wrong one (i.\,e. the one corresponding to the other context signal), both before turn-off and after it, are provided in Fig.\ref{fig2}D for both the neural architectures.

Furthermore, apical context architecture is also robust against corruption in the context signal, which may be the case when at higher cortex level there is only a mild preference in favor of which strategy to adopt (in comparison with the training phase, where each target is clearly and univocally associated with a sharp context signal). Here a Gaussian white noise of variance $\sigma^2$ is added during test to context signals exploited in the training (Fig.\ref{fig2}C, top panel, $\sigma=0.2$). The produced trajectory is similar to the trajecory targeted by the context A (black dashed line) and different from the trajectory targeted by the context B.
In Fig.\ref{fig2}C, bottom panel (blue lines) it is reported the average $mse$ (average over 10 independent realizations of the experiment) between the output and the target trajectory (solid blue line) as a function of $\sigma$. As a reference, we also report the $mse$ between the output and the trajectory targeted by the other context signal (dashed blue line).

It is evident a resilience of the network with apical context, while the network with basal context suddenly loses the ability to reproduce the desired output already at low levels of noise (Fig.\ref{fig2}C, top panel and bottom panel orange lines). 

At higher level of noise, basal-context network becomes in practice useless, while apical-context network is still able to reproduce the target trajectory with a remarkably small error (Fig.\ref{fig2}C, bottom panel).

\subsection{Hierarchical Imitation Learning}

\begin{figure*}[h!]
\centering
\includegraphics[width=160mm]{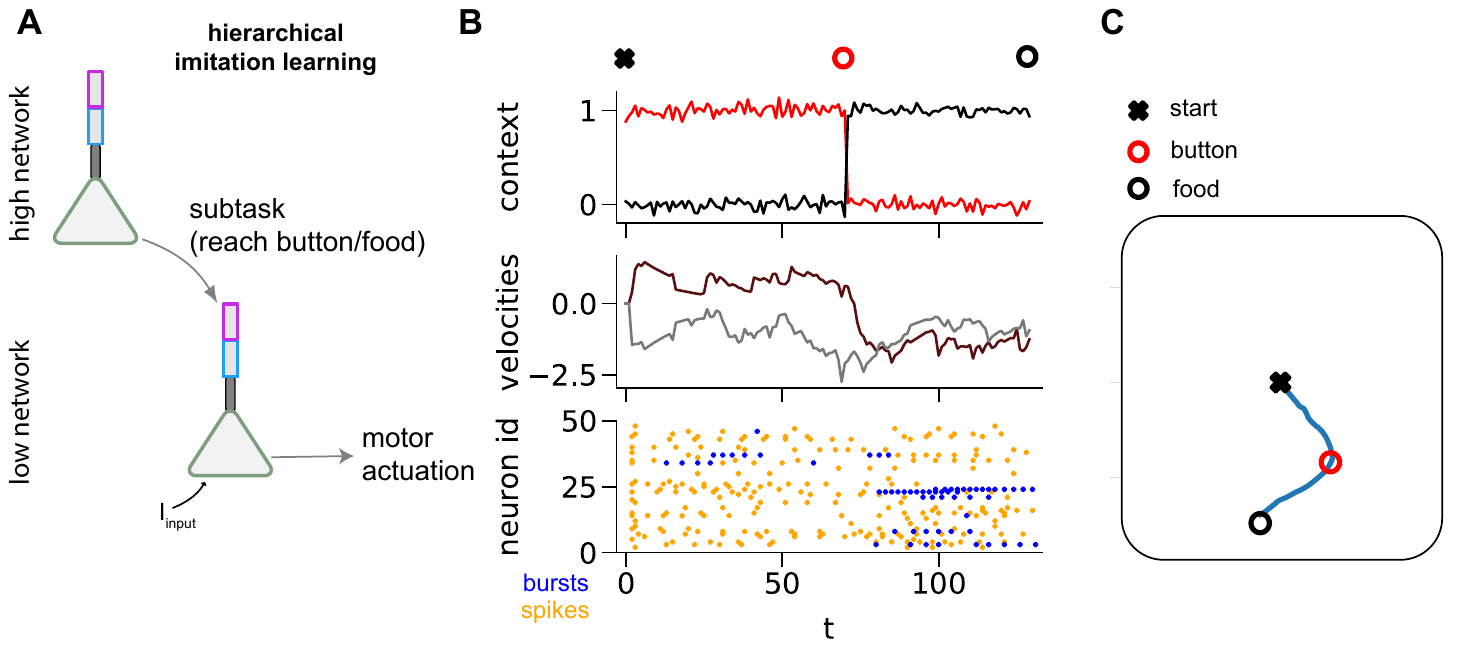}
\caption{
\textbf{Hierarchical Imitation Learning} 
\textbf{A}. A two-level network, where high-level neurons produce a signal that serves as a context for the neurons in the low-level network. The two subnetworks received two different but synchronized teaching signals in the training phase.
\textbf{B}. Button-and-food task, an agent placed at an initial position (black cross) in a 2D maze has to first, reach a button (red circle) and then the food (black circle). The high-level network chooses the order of the subtasks: reach-the-button and then reach-the-food, and projects the instruction as a contextual signal (top panel) to the apical compartments of the low-level network. The low-level network produces the output (velocities of the agent, center panel) necessary to solve the subtask as a readout of its internal bursting activity (bottom panel, blue dots). Orange dots represent the spiking activity. 
\textbf{C}. A sample spatial trajectory. Cross, red and black circles as in panel B.
}
    \label{fig3}
\end{figure*}

The proof that context can be used to flexibly  choose which dynamics reproduce (and when), opens the pathway to more complicated neural architectures, naturally supporting hierarchical imitation learning. To our knowledge, no prior works are proposing biologically plausible implementations of hierarchical reinforcement or imitation learning.

We decomposed the network in two sub-networks which we call high-network and low-network (Fig.\ref{fig3}A). The high-network (manager) computes the optimal strategy to take to solve a task and sends this information as a context signal to the low-network (worker) which actually executes it.

We applied this strategy to the so-called \emph{button \& food} task. In this task, an agent starts at the center of a square domain, which also features a button and an initially locked target (the food). The goal of the agent is to first press the button so to unlock the food and then reach for it. Both button and food positions are uniformly extracted in the domain $\left[0, 1 \right] \times \left[0, 1 \right]$. 
The global task is naturally decomposed into two sub-tasks (or goals): $ \mathsf{reach\_button}$ and $\mathsf{reach\_food}$. The high-network computes which goal to pursue and when, and the low-network implements the sub-policy to achieve the goal.

Both the high- and the low-network share the same input, ( $I = 80$ input units) the vertical and horizontal differences of both the button's and food's positions with respect to agent location ($\Delta^t =\{\Delta x_b^t, \Delta y_b^t,\Delta x_f^t, \Delta y_f^t\}$ respectively). These quantities are encoded through a set of tuning curves. Each of the $\Delta_i$ values are encoded by 20 input units with different Gaussian activation functions.


To perform learning, we consider a natural hierarchical extension of behavioral cloning. The expert provides a set of hierarchical demonstrations, each consisting of low-level trajectories (to be cloned by the low-network):
\begin{equation*}
\left\{ \left( \mathrm{state}^t_\mathrm{L}, \mathrm{action}^t_\mathrm{L}, \mathrm{goal}^t_\mathrm{L} \right)\right\}_{t=1}^T,
\end{equation*}
as well as a high-level trajectory (to be cloned by the high-network):
\begin{equation*}
\{\left( \mathrm{state}^t_\mathrm{H}, \mathrm{action}^t_\mathrm{H} \right)\}_{t=1}^T.
\end{equation*}

Both $\mathrm{state}^t_\mathrm{L}$ and $ \mathrm{state}^t_\mathrm{H}$ are the input $\Delta^t$ described above.

The $\mathrm{action}^t_\mathrm{H}$ is the target output of the high-network and the $\mathrm{goal}^t_\mathrm{L}$ of the low-network. It is projected to the low-network as a contextual signal in the distal apical compartment (Fig.\ref{fig3}B, top) and is defined as a binary two-dimensional teaching signal: 
\begin{equation*}
\bm{y}^{\star, t}_\mathrm{H} = \bm{\chi}_{(1)} \Theta \left(t < t_b \right) + \bm{\chi}_{(2)} \Theta \left(t > t_b \right),
\end{equation*}
where $\bm{\chi}_{(i)}$ is one for vector-components $i$-th and zero otherwise and $t_b$ is the time when the button is reached. Intuitively, this target selects the $\mathsf{reach\_button}$ sub-policy for the first part of the task and then switches to $\mathsf{reach\_target}$.

Given the input $\mathrm{state}^t_\mathrm{L}$ and the context $\mathrm{goal}^t_\mathrm{L}$, the low-network 
is tasked to produce as output $\mathrm{action}^t_\mathrm{L}$, the velocity vector $\bm{y}^{\star, t}_\mathrm{L} = \bm{v}^t = (v_x^t, v_y^t)$, where the velocities are computed so to reach the selected target in a straight line (Fig.\ref{fig3}B, center and Fig.\ref{fig3}C) and the output is computed as a linear readout of its internal bursting activity (Fig.\ref{fig3}B, bottom). 

The cloning procedure is implemented as a supervised learning to make the two networks reproduce the target outputs, given the input (and the context). The learning procedure is the same as the one described in Section 2.1. Finally, the two layer network is tested in closed-loop in the environment described above.

The performance in this task is measured via the following quantity:

\begin{equation*}
    \rho = \frac{\Xi_{\lock}}{\min_t d \left(\bm{x}^t_\mathrm{agent}, \bm{x}_\mathrm{food} \right)},
\end{equation*}

where $\Xi_{\lock}$ is the button-state indicator variable that is zero when the button is locked and one otherwise, the $\bm{x}^t_{(\cdot)}$ are the agent and target position vectors and $d \left( \cdot, \cdot \right)$ is the standard euclidean distance. The condition for a successful button-press (a switch between locked and unlocked) and target-reach is taken to be $d \left(\bm{x}^t_\mathrm{agent}, \bm{x}_\mathrm{btn|food} \right) \le 0.1$. Note how effectively this choice prevents the apparent divergence in the expression for $\rho$ as the episode is stopped when the target is reached, which induces a theoretical maximum achievable score of $\rho_\mathrm{max} = 10$.  


After the presentation of many randomly positioned button-food pairs, we observe that such two-level network learns to correctly and efficiently solve the button \& food task, with an average final score $\rho = 7.3 \pm 4.8$ and over $90\%$ of success rate (i.e. both button-press and target-reach conditions were met). A sample spatial trajectory produced by the network is depicted in Fig.\ref{fig3}C.

\section{Methods}

\subsection{The model}

Our model of pyramidal neuron considers three different compartments: a basal one ($b$) and two apical ones, named proximal ($p$) and distal ($d$), respectively (see Figure \ref{fig1} for reference). 

Consider a particular neuron $i$, with $i = 1, \dots, N$,  its real vector-valued membrane potential $\bm{\mathrm{v}}^t_i = \left(v^t_i, u^t_i, u_i^{\star, t} \right)$ (the membrane potentials of the basal, the proximal apical, and distal apical compartments respectively) follows a leaky integrate and fire dynamics, which we can generically write as:
\begin{equation}
    \bm{\mathrm{v}}_i^{t+1} = \left[\left(1 - \frac{dt}{\tau_m} \right) \bm{\mathrm{v}}_i^t + \frac{dt}{\tau_m} \bm{\mathrm{I}}_i^{t+1} \right] \left(1 - \bm{\mathrm{s}}_i^t \right) + \bm{\mathrm{v}^{\circlearrowright}}  \bm{\mathrm{s}}_i^t
\label{EQ:Vector_Membrane_Potential_Dynamics}
\end{equation}
where the vector valued quantities $\bm{\mathrm{I}}^t_i = (I_{(b), i}^t, I_{(p), i}^t, I_{(d), i}^t )$ $\bm{\mathrm{s}}_i^t = (z_i^t, a_i^t, a_i^{\star, t})$ and $\bm{\mathrm{v}^{\circlearrowright}} = (\mathrm{v}^\circlearrowright_{(b)}, \mathrm{v}^\circlearrowright_{(p)}, \mathrm{v}^\circlearrowright_{(d)})$, respectively the input current, the neuron spike and the reset potential, depend on the compartment (see following sections for explicit definitions). In particular, the neural spike $\bm{\mathrm{s}}_i^t$ is a stochastic variable determined via its sigmoidal probability:
\begin{equation}
    p \left(\bm{\mathrm{s}}_i^{t+1} | \bm{\mathrm{v}}_i^t \right) = \frac{\exp \left[\bm{\mathrm{s}}_i^{t+1} \left(\frac{\bm{\mathrm{v}}_i^t - v_\mathrm{thr}}{\delta v} \right) \right]}{1 + \exp \left(\frac{\bm{\mathrm{v}}_i^t - v_\mathrm{thr}}{\delta v} \right)}
\label{EQ:Vector_Probability_Sigmoid_Definition}
\end{equation}
with $v_{\mathrm{thr}}$ being the firing threshold for the membrane potential and $\delta v$ a model parameter controlling the probabilistic nature of the neuron. In the $\delta v \to 0$ limit, the spike-generation rule \eqref{EQ:Vector_Probability_Sigmoid_Definition} becomes deterministic:
\begin{equation*}
p(\bm{\mathrm{s}}^{t+1}|\bm{\mathrm{v}}^t)=\Theta[\bm{\mathrm{s}}^{t+1}(\bm{\mathrm{v}}^t-v_{\mathrm{thr}})]. 
\end{equation*}
We remark that we assume the deterministic limit to numerically implement the dynamics ($\delta v\rightarrow0$).
\subsubsection{Temporal filtering and windows}
We introduce the exponential filtering function $\mathsf{filter}\left(\xi^t, \tau \right)$, defined recursively as:
\begin{multline}
    \mathsf{filter} \left( \xi^{t+1}, \tau \right) = \exp \left(-\frac{dt}{\tau} \right) \mathsf{filter} \left( \xi^t, \tau \right) +\\+ \left( 1 - \exp \left(-\frac{dt}{\tau} \right) \right) \xi^{t+1}.
\label{EQ:Filtering_Function}
\end{multline}
Basal spike signals are time-filtered through suitable time constants, depending on the direction they propagate. Using the previous definition, we introduce the following filtered quantities:
\begin{align}
\hat{z}_{i}^{t+1} &= \mathsf{filter} \left(z_{i}^{t+1}, \tau_s \right) \\
\hat{z}_{\mathrm{ro},i}^{t+1} &= \mathsf{filter} \left(z_{i}^{t+1}, \tau_\mathrm{ro} \right) \\
\hat{z}_{\mathrm{soma},i}^{t+1} &= \mathsf{filter} \left(z_{i}^{t+1}, \tau_\mathrm{targ} \right)
\end{align}
Such filtering is also applied to the adaptatoin term $\omega_i^t$, which is time-smoothed as:
\begin{equation}
    \omega_i^{t+1} = \mathsf{filter} \left(z_{i}^{t+1}, \tau_\omega \right).
\end{equation}
Coincidence between above-threshold somatic spikes $\hat{z}_{\mathrm{soma}, i}^t$ and apical proximal $a_i^t$ or apical distal $a_i^{\star, t}$ spikes opens a time-window $\shat{z}_i^t$ for bursts onset $\bm{\mathrm{B}}_i^t = (B_{i}^t, B_{i}^{\star, t})$:
\begin{equation}
\shat{z}_{i}^{t} = \Theta[\hat{z}_{\mathrm{soma},i}^{t} - \vartheta_{\mathrm{soma}}]
\end{equation}
The burst variable $\bm{\mathrm{B}}_i^t \in \left\{0, 1 \right\} \times \left\{0, 1 \right\}$ is a tuple of binary variables signaling the onset of a burst activity in the proximal or distal compartments, which can be expressed as:
\begin{align}
B_{i}^{t+1} &= \shat{z}_i^{t}  a_{i}^{t+1}\\
B_{i}^{\star, t+1} &= \shat{z}_i^{t} a_{i}^{\star, t+1}
\end{align}
Aiming for a time-window variable that is active during burst activity, we can iterate the same construction developed for spikes and consider the filtered burst-onset $\hat{\bm{\mathrm{B}}}_i^t$:
\begin{align}
\hat{B}_{i}^{t+1} &= \mathsf{filter} \left( B_i^{t+1}, \tau_\mathrm{targ} \right)\\
\hat{B}_{i}^{\star,t+1} &= \mathsf{filter} \left( B_i^{\star, t+1}, \tau_\mathrm{targ} \right)
\end{align}
One can again use this filtered quantities to introduce proximal and distal burst windows as:
\begin{align}
\shat{B}_{i}^{t+1} &= \Theta[\hat{B}_{i}^{t+1} - \vartheta_{\mathrm{burst}}]\\
\shat{B}_{i}^{\star, t+1} &= \Theta[\hat{B}_{i}^{\star, t+1} - \vartheta_{\mathrm{burst}}]
\end{align}
When at least one among proximal and distal bursts is above threshold, we finally have a neural burst activity window:
\begin{equation}
    \shat{B}_{\lor,i}^{t+1} = \shat{B}_{i}^{t+1} \lor \shat{B}_{i}^{\star, t+1},
\end{equation}
which is the quantity that will feature in the dynamics of the compartments.
\subsubsection{Basal compartment}
The membrane potential of the basal compartment evolves following the equations:
\begin{align*}
v_{i}^{t+1} &= \left[\left(1-\frac{dt}{\tau_m}\right)v_{i}^t + \frac{dt}{\tau_m}I_{(b),i}^{t+1}\right](1 - z_{i}^t) + \mathrm{v}^{\circlearrowright}_{(b)} z_{i}^t \\
I_{(b),i}^t &= \sum_{j=1}^N J_{ij}^{b\to b} \hat{z}_{j}^t + \sum_{k=1}^{n_{\mathrm{inp}}} J_{ik}^{\mathrm{inp}} I_k^{\mathrm{inp},t} + \beta \shat{B}_{\lor, i}^t - b \hat{\omega}_i^t + v_0
\end{align*}
With $J_{ik}^\mathrm{inp}$ and $I_k^{\mathrm{inp}, t}$ respectively the input connection matrix and current, while $v_0$ is a compartment-specific constant input.
We introduced the basal reset potential:
\begin{equation*}
    \bm{\mathrm{v}}^{\circlearrowright}_{(b)} = \frac{v_{\mathrm{reset}, b}}{1 + \alpha \shat{B}_{\lor, i}^{t}},
\end{equation*}
Where $v_{\mathrm{reset}, b}$ is a compartment-specific scalar, $\alpha$ is a constant model parameter and $\shat{B}_{\lor, i}^t$ is the active burst-window variable (see section \textsc{temporal filtering and windows} for and explicit characterization).
Note how during the burst-window $\shat{B}_{\lor,i}^t$ the soma receives an extra input and the reset potential is higher, we set $\alpha=2$ and $\beta=20$ to define the entity of such effects.
\subsubsection{Apical proximal compartment}
The apical proximal compartment of each neuron is connected to basal compartments of all the neurons through recurrent connections $J_{ij}^{b \to p}$ (the ones to be trained to reproduce the desired target).  The equation for this compartment's dynamics are:
\begin{align*}
u_{i}^{t+1} &= \left[\left(1-\frac{dt}{\tau_m}\right) u_{i}^t + \frac{dt}{\tau_m}I_{(p),i}^{t+1}\right](1 - a_{i}^t) + \mathrm{v}^\circlearrowright_{(p)} a_{i}^t\\
I_{(p),i}^t &= \underbrace{\sum_{j=1}^N J_{ij}^{b \to p} \hat{z}_{j}^t (t)}_{\substack{\mathsf{recurrent\   basal-proximal}\\ \mathsf{connections}}} +\ u_0
\end{align*}
The reset potential for the proximal apical compartment $\mathrm{v}^\circlearrowright_{(p)} = v_{\mathrm{reset}, p}$ is a compartment-specific scalar, independent of burst activity, while $u_0$ is the compartment constant input.
\subsubsection{Apical distal compartment}
The signal to be learned (target) is considered as an input for the apical distal compartment: coefficient $f_{\mathrm{apic}}$ is set to $1$ during the learning stage, and then set to $0$ to get rid of this term during spontaneous activity. Also, the input from the context (again randomly projected on the $N$ neurons) is given as input for the apical distal compartment. The equations for the apical distal compartment read:
\begin{align*}
u_{i}^{\star, t+1} &= \left[\left(1-\frac{dt}{\tau_m}\right) u_{i}^{\star, t} + \frac{dt}{\tau_m} I_{(d),i}^{t+1}\right](1 - a_{i}^{\star, t}) + \mathrm{v}^{\circlearrowright}_{(d)} a_{i}^{\star, t} \\
I_{(d),i}^t &= \underbrace{f_\mathrm{apic}\sum_{k=1}^{n_{\mathrm{output}}} J^{\mathrm{targ}}_{ik}y_k^{\star, t}}_{\mathsf{target/teach\ input}} + \underbrace{\sum_{k=1}^{n_{\mathrm{cont}}}J_{ik}^{\mathrm{cont}}C_k^t}_{\mathsf{context}} +\  u^\star_0 
\end{align*}
where $y_k^{\star, t}$ is the target signal and $C_k^t$ the context signal, while $u_0^\star$ is the compartment constant input.
We report the model parameters, for the three figures, in Table.\ref{table1}.
\begin{table}[t]
\caption{\textbf{Parameter of numerical simulations}. Many parameters have the same value for all the simulations reported in the main text figures. When not the case, the different values used are clearly indicated. For \textsc{fig} \ref{fig3} two values for low network (L) and high network (H), respectively, have been reported, when different from each other. For \textsc{fig} \ref{fig2} $\eta$ and $\eta_{\mathrm{out}}$ we report the initial parameter values, during learning they are discounted as discussed in section 2.2.}
\label{table1}
\vskip 0.15in
\begin{center}
\begin{small}
\begin{sc}
\begin{tabular}{lcccr}
\toprule
Parameter & Fig 1 & Fig 2 & Fig 3 [L\,--\,H]\\
\midrule
$N$                             &  500   &  1000  &  500--500     \\
$\sigma_{\mathrm{targ}}$       &  20    &  30    &  0\,--\,100   \\
$\sigma_{\mathrm{in}}$          &  12    &  12    &  20           \\
$\eta$                          &  10     &  10    &  0\,--\,0.25  \\
$\eta_{\mathrm{out}}$           &  0.01  &  0.01  &  0.03         \\
$I$                             &  5     &  50    &  n.d.         \\
$\sigma_{\mathrm{cont}}$        &  0     &  20    &  50\,--\,0    \\
\midrule
$N_{e}$                         &  \multicolumn{3}{c}{$80\%\,N$}             \\
$N_{i}$                         &  \multicolumn{3}{c}{$20\%\,N$}             \\
$\tau_{\mathrm{m}}$             &  \multicolumn{3}{c}{20 $\mathrm{(ms)}$}    \\
$\tau_{\mathrm{s}}$             &  \multicolumn{3}{c}{2 $\mathrm{(ms)}$}     \\
$\tau_{\mathrm{out}}$           &  \multicolumn{3}{c}{10 $\mathrm{(ms)}$}    \\
$\tau_{\mathrm{targ}}$          &  \multicolumn{3}{c}{20 $\mathrm{(ms)}$}    \\
$\tau_{\omega}$                 &  \multicolumn{3}{c}{200 $\mathrm{(ms)}$}   \\
$b$                             &  \multicolumn{3}{c}{100}                   \\
$v_{\mathrm{reset}, b}$         &  \multicolumn{3}{c}{-20 $\mathrm{(mV)}$}   \\
$v_{\mathrm{reset}, d, p}$      &  \multicolumn{3}{c}{-160 $\mathrm{(mV)}$}  \\
$v_0$                           &  \multicolumn{3}{c}{-1 $\mathrm{(mV)}$}    \\
$u_0$                           &  \multicolumn{3}{c}{-6 $\mathrm{(mV)}$}    \\
$u_0^{\star}$                   &  \multicolumn{3}{c}{-6 $\mathrm{(mV)}$}    \\
$v_\mathrm{thr}$                &  \multicolumn{3}{c}{0 $\mathrm{(mV)}$}     \\
$\vartheta_{\mathrm{soma}}$     &  \multicolumn{3}{c}{$2.5 \times 10^{-2}$}  \\
$\vartheta_{\mathrm{burst}}$    &  \multicolumn{3}{c}{$1.25 \times 10^{-2}$} \\
\bottomrule
\end{tabular}
\end{sc}
\end{small}
\end{center}
\vskip -0.1in
\end{table}

\subsection{Derivation of the learning rule}
We derive the update rule for the recurrent weights of the network by maximizing the probability to reproduce the target spatio-tamporal pattern of bursts, extending previous approaches used for learning target pattern of spikes \cite{pfister2006optimal,rezende2014stochastic,gardner2016supervised,muratore2021target}.
The first step is to write the probability to produce a burst in the neuron $i$ at time $t$, given the somatic window $\shat{z}_i^t$. We propose the following compact formulation:
\begin{equation}
    p(B_{i}^{\star, t+1}| \shat{z}_{i}^t) = \frac{\exp{\left[B_{i}^{\star, t+1} \Phi_{i}^t (\shat{z}_{i}^t) \right]}}{1+\exp{\left[ \Phi_i^t(\shat{z}_{i}^t) \right] } }
\end{equation}
where we have introduced $\Phi_{i}^t (\shat{z}_{i}^t) = a_i^t \shat{z}_{i}^t/\delta v - (1-\shat{z}_{i}^t)\gamma$. By definition, a burst can only happen by means of a basal-apical spike coincidence, represented by the $a_i^t \shat{z}_i^t$ term. When the basal window is open ($\shat{z}_i^t = 1$) the burst probability reduces to the usual sigmoidal function. When the window is closed and $\shat{z}_i^t = 0$, we have $\Phi_i^t \left( \shat{z}_i^t \right) = -\gamma$, we can thus tune the $\gamma$ parameter to model the burst probability. In practice, we work in the $\gamma \to \infty$ limit where $\lim_{\gamma \to \infty} p (B_i^{\star, t + 1} | \shat{z}_i^t = 0) = 0$, which agrees to the intuitive understanding that a closed basal window prevents any burst activity.  
We introduce the likelihood $\mathcal{L}$ of observing a given target burst activity $\bm{\mathrm{B}^\star}$ given the basal-to-proximal connections $J_{ij}^{b \to p}$ as:
\begin{multline}
\mathcal{L} \left( \bm{\mathrm{B}^{\star}} | J^{b \to p} \right) = \sum_{it} \left[ B_{i}^{\star, t+1} \Phi_{i}^t (\shat{z}_{i}^t) \right. +\\- \log{ \left( 1 + \exp{\left[ \Phi_i^t(\shat{z}_{i}^t) \right] } \right)} \Big]
\end{multline}
We can then maximize this likelihood by adjusting the synaptic connection so to achieve the target burst activity $\bm{\mathrm{B}^\star}$. By differentiating with respect to the recurrent apical weights, we get:
\begin{equation}
\frac{\partial \mathcal{L}( \bm{\mathrm{B}}^{\star} | J^{b \to p})}{\partial J_{ij}^{b \to p}} =  \left[B_{i}^{\star, t+1} -p( B_{i}^{t+1}=1) \right] \shat{z}_{i}^t e_j^t
\end{equation}
where we have introduced the following two quantities:
\begin{equation*}
p( B_{i}^{t+1}=1) =\frac{\exp{\left[ \Phi_{i}^t (\shat{z}_{i}^t) \right]}} {1+\exp{\left[ \Phi_i^t(\shat{z}_{i}^t) \right]}} \quad \mathrm{and} \quad e_j^t = \frac{\partial u_i^t}{ \partial J_{ij}^{b \to p}}.
\end{equation*}
Given the basal window $\shat{z}_i^t$ state, the target burst sequence is uniquely defined by the input projected to the apical distal compartment and can be written as $B_i^{\star, t+1} = \shat{z}_i^t a_{i}^{\star, t+1}$.
If we take the model deterministic limit ($\delta v \rightarrow 0$, where $p (B_i^{t+1} = 1) = a_i^{t+1} \shat{z}_i^t)$ and note that $\shat{z}_i^t \shat{z}_i^t = \shat{z}_i^t$, we can rewrite the previous expression in a cleaner form:
\begin{equation}
\frac{\partial \mathcal{L} ( \bm{\mathrm{B}^\star}  | J^{b \to p})}{\partial J_{ij}^{b \to p}} =  \left[a_{i}^{\star, t+1} - a_{i}^{t+1} \right] \shat{z}_{i}^t e_j^t.
\end{equation}
This means that the spikes in the proximal apical compartment $a_{i}^{t+1}$ should mimic the ones in the distal one $a_{i}^{\star, t+1}$, when the somatic window $\shat{z}_i^t$ is open.
For simplicity, we discussed this version of the learning rule. However, in this work we used the non-deterministic version of the rule (finite $\delta v = 0.1$)
that can be rewritten as:
\begin{equation}
\frac{\partial \mathcal{L} ( \bm{\mathrm{B}^\star}  | J^{b \to p})}{\partial J_{ij}^{b \to p}} =  \left[a_{i}^{\star, t+1} - p \left(\mathrm{a}_i^{t+1}=1 | \mathrm{u}_i^t \right) \right] \shat{z}_{i}^t e_j^t.
\end{equation}
where $p \left(\mathrm{a}_i^{t+1}=1 | \mathrm{u}_i^t \right) = \frac{\exp   \left(\frac{\mathrm{u}_i^t - v_\mathrm{thr}}{\delta v} \right)  }{1 + \exp \left(\frac{\mathrm{u}_i^t - v_\mathrm{thr}}{\delta v} \right)}$.
We stress here how in the derivation we considered the basal-windows state $\shat{z}_i^t$ as given. Consequently, the target burst sequence $\bm{\mathrm{B}^\star}$ is uniquely defined by the input projected to the apical distal compartment and the likelihood is well defined. We are aware however of the feedback influence of the burst activity on the basal-window configuration (bursts induce basal spikes, see the equation for basal current $I_{(b), i}^t$ in the \textsc{basal compartment} section), we chose to neglect such contribution as it would have severely increased the difficulty of the derivation. The convergence to the chosen target thus cannot be granted. Despite the fact that we cannot theoretically prove the convergence of the learning rule, we provide a numerical demonstration that the target pattern of bursts converges to a well defined pattern (see Appendix for details).

\subsection{Source code availability}
The source code is available for download under CC-BY license in the
\\ \url{https://github.com/cristianocapone/LTTB} public repository.

\section{Discussion}
In the present work, we have shown that the anatomy of pyramidal neuron can naturally support target-based learning. Moreover, it allows for using contextual signals to flexibly select the desired output from a repertoire of learned dynamics. 

These properties naturally combine together to orchestrate a network with a hierarchical architecture, which in turn lends itself to \emph{hierarchical imitation learning} (HIL) \cite{le2018hierarchical}. HIL enables the decomposition of challenging long-horizon decision-making tasks into simpler sub-tasks, improving both learning speed and transfer learning, as skills learned by sub-modules can be re-used for different tasks. In our work, a high-level network (the manager) selects the correct policy for the task, while the low-level network (the worker) is in charge of actually executing it. 


To our knowledge, there exist no other works proposing a biologically plausible architecture to implement HIL.
Furthermore, our model prepares the ground for further biological explorations. Model parameters (e.g., the adaptation strength $b$) allows simulating the transition between different brain states (e.g., sleep and awake) \cite{wei2018differential,goldman2020brain,tort2021attractor}. Possible future investigation topics include replay of the pattern of bursts during sleep \cite{kaefer2020replay}, and the effect of sleep on tasks performances \cite{wei2018differential,capone2019sleep}.



\section*{Acknowledgement}
This work has been supported by the European Union Horizon 2020 Research and Innovation program under the FET Flagship Human Brain Project (grant agreement SGA3 n. 945539 and grant agreement SGA2 n. 785907) and by the INFN APE Parallel/Distributed Computing laboratory.

\bibliography{references.bib}
\bibliographystyle{apalike}

\newpage
\appendix

\onecolumn

\begin{center}
\Large \textbf{Appendix: Burst-dependent plasticity and dendritic amplification support target-based
learning and hierarchical imitation learning}
\end{center}

\section{Numerical evidence of convergence}

As mentioned above, we can not provide a mathematical proof of the convergence toward the chosen target of burst activity by means of the learning rule proposed here. However, strong evidences in this direction can be found numerically.

We run several independent realizations of the same task of Fig.\ref{fig1}, i.e., the store-and-recall of a 3D trajectory. We look at the distance between the target and the spontaneous spatio-temporal pattern of bursts during the training, and also at the self-distance in the pattern of spontaneous bursts across consecutive training iterations.

The parameters used for these simulations (when different from those used for Fig.\ref{fig1}) are: $\eta=2.5$, $\eta_{\mathrm{out}}=2.5 \times 10^{-3}$, $\sigma_{\mathrm{targ}}$ variable from $10$ (black) to $1000$ 
(yellow). Data averaged over $10$ independent network/target realizations.
The distance between two patterns of bursts $A =\{A_i^t\}$ and $B = \{B_i^t\}$ is defined as:

\begin{equation*}
    \mathcal{D}(A,B) \equiv \sqrt{\frac{1}{N\,T}\sum_{i=1}^N \sum_{t=1}^T \left(A^t_i-B_i^t\right)^2}
\end{equation*}

\begin{figure}[h]
\centering
\includegraphics[width=16cm]{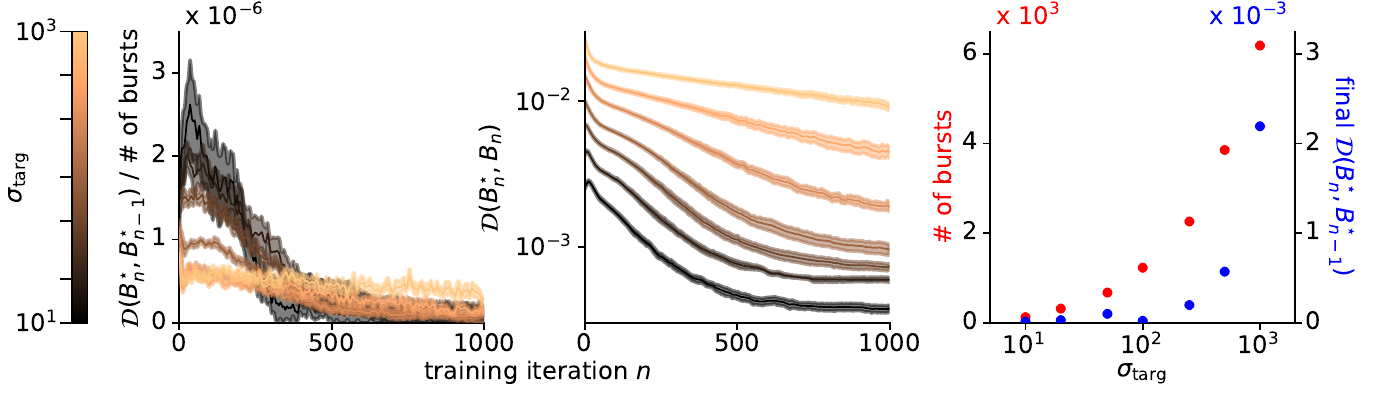}
\caption{
\textbf{Convergence of the target pattern of bursts.}
\textbf{(left)}  $\mathcal{D}(B^{\star}_n,B^{\star}_{n-1})/(number\,of\,bursts)$  as a function of the number $n$ of learning iterations, for different  $\sigma_{\mathrm{targ}}$ values (lower to higher values, from dark to light). \textbf{(middle)}  Distance between the target and spontaneous pattern of bursts $\mathcal{D}( B^{\star}_n,B^{n} )$ after $n$ learning iterations. \textbf{(right)} Blue: average final $\mathcal{D}(B^{\star}_n,B^{\star}_{n-1})/(number \, of \, bursts)$ value as a function of $\sigma_{\mathrm{targ}}$. Red: average $number \, of \, bursts$ as a function of $\sigma_{\mathrm{targ}}$.
}
\label{figs1}
\end{figure}

For small values of $\sigma_{\mathrm{targ}}$, comparable to the ones used for main text figures, target bursts rapidly settle after some hundreds of training iterations  (Fig.\ref{figs1}A); within the same training scale, also spontaneous burst activity matches the target one, with a negligible error (Fig.\ref{figs1}B). Accordingly, the overall number of bursts is the same for target and spontaneous activity (Fig.\ref{figs1}C).

We prove that in a broad range of $\sigma_{\mathrm{targ}}$ values, the target pattern of bursts converges to a well defined one ((Fig.\ref{figs1}C blue dots), up to $\sigma_{\mathrm{targ}}=100$ ) even though the number of bursts increases for high values of $\sigma_{\mathrm{targ}}$ (Fig.\ref{figs1}C red dots).

\end{document}